\documentclass[sigconf]{acmart}
\acmConference[EASE 2025]{The 29th International Conference on Evaluation and Assessment in Software Engineering}{17–20 June, 2025}{Istanbul, Türkiye}

\PassOptionsToPackage{table,xcdraw}{xcolor}
\usepackage{booktabs} 
\usepackage{multirow,xspace,pifont}
\usepackage{xcolor}
\usepackage{colortbl}
\usepackage{amsmath}
\usepackage[ruled,linesnumbered]{algorithm2e}
\usepackage{graphicx}
\usepackage{textcomp}
\usepackage{tcolorbox}
\usepackage{siunitx}
\usepackage{mdframed}

\usepackage{listings,verbatim}
\usepackage{caption}
\usepackage{subcaption}
\usepackage{cleveref}

\AtBeginDocument{%
  }

\setcopyright{acmlicensed}
\copyrightyear{2025}
\acmYear{2025}
\acmDOI{XXXXXXX.XXXXXXX}

\acmISBN{978-1-4503-XXXX-X/18/06}

\begin{document}

\title{Accelerating Delta Debugging through Probabilistic Monotonicity Assessment}

\author{Yonggang Tao}
\orcid{0009-0006-3669-2047}
\affiliation{%
  \institution{University of New South Wales}
  \city{Sydney}
  \state{NSW}
  \country{Australia}
}
\email{yonggang.tao@unsw.edu.au}

\author{Jingling Xue}
\orcid{0000-0003-0380-3506}
\affiliation{%
  \institution{University of New South Wales}
  \city{Sydney}
  \state{NSW}
  \country{Australia}
}
\email{j.xue@unsw.edu.au}

\renewcommand{\shortauthors}{Tao et al.}

\newcommand{\PMA}{\textsc{PMA}\xspace}
\newcommand{\CHISEL}{\textsc{CHISEL}\xspace}
\newcommand{\ProbDD}{\textsc{ProbDD}\xspace}
\newcommand{\ddmin}{\textsc{ddmin}\xspace}
\newcommand{\true}{\textbf{true}\xspace}
\newcommand{\false}{\textbf{false}\xspace}
\newcommand{\confun}{\mbox{$\mathfrak{C}$}\xspace}
\newcommand{\monoProp}{\textsc{MonoUpdate}\xspace}
\newcommand{\monoUpdate}{\textsc{MonoUpdate}\xspace}
\newcommand{\monoObserver}{\textsc{monoAssr}\xspace}
\newcommand{\monoCheck}{\textsc{SkipEnabled}\xspace}
\newcommand{\skipTest}{\textsc{SkipAllowed}\xspace}
\newcommand{\monoCount}{\mathcal{M}\xspace}
\newcommand{\supTest}{\textsc{SupSet}\xspace}

\newcommand{\cmark}{\ding{51}}
\newcommand{\xmark}{\ding{55}}

\newtheorem{assumption}{Assumption}

\begin{abstract}

Delta debugging assumes search space monotonicity: if a program causes a failure, any supersets of that program will also induce the same failure, permitting the exclusion of subsets of non-failure-inducing programs. However, this assumption does not always hold in practice. This paper introduces Probabilistic Monotonicity Assessment (\PMA), enhancing the efficiency of \ddmin-style algorithms without sacrificing effectiveness. \PMA dynamically models and assesses the search space's monotonicity based on prior tests tried during the debugging process and uses a confidence function to quantify monotonicity, thereby enabling the probabilistic exclusion of subsets of non-failure-inducing programs. Our approach significantly reduces redundant tests that would otherwise be performed, without compromising the quality of the reduction.

We evaluated \PMA against two leading \ddmin-style tools, \CHISEL and \ProbDD. Our findings indicate that \PMA cuts processing time by 59.2\% compared to \CHISEL, accelerates the reduction process (i.e., the number of tokens deleted per second) by $3.32\times$, and decreases the sizes of the final reduced programs by 6.7\%. Against \ProbDD, \PMA reduces processing time by 22.0\%, achieves a $1.34\times$ speedup in the reduction process, and further decreases the sizes of the final reduced programs by 3.0\%. These findings affirm \PMA's role in significantly improving delta debugging’s efficiency while maintaining or enhancing its effectiveness.
\end{abstract}



\maketitle

\section{Introduction}
\label{sec:intro}


The scale and complexity of software have rapidly increased, raising the risk of severe bugs. For example, an update from CrowdStrike on July 19, 2024 caused 8.5 million Windows devices to crash, resulting in over one billion dollars in losses \cite{Weston24}. This underscores the critical need for efficient bug identification. Manual bug identification in large codebases is time-consuming \cite{csmith}, with software maintenance consuming nearly 70\% of development resources, primarily for bug localization and repair \cite{maintenancecost}.

\emph{Delta debugging} \cite{zeller1999} was introduced to minimize failure-inducing programs while maintaining the failure. This approach has been widely applied across various software engineering domains, including compiler testing \cite{csmith, emi, livecode, chen2019history, 2021test}, SMT solver testing \cite{yinyang, 2021skeletal}, fault localization \cite{zellerlocating, zellerisolating}, and software debloating \cite{zhang2022one, chisel}.
 
Formally, delta debugging is defined as follows: Let \( S \) be a program, represented as a set of elements or tokens. Let \( \mathbb{S} = 2^S \) denote the search space, which includes all subsets of \( S \), referred to as test cases \( S' \). The property test function \( \psi: \mathbb{S} \to \{\true, \false\}\) evaluates each test case \( S' \),
returning \(\true\) if \( S' \) maintains the failure-inducing property and \false otherwise. Given \( \psi(S) = \true \), delta debugging aims to efficiently find the smallest subset \( S^* \in \mathbb{S} \) that still maintains this property (\( \psi(S^*) = \true \)), thus minimizing the object's size while preserving the failure-inducing property.


In compiler testing, failure-inducing test cases are often large. Both GCC and LLVM recommend test case reduction in bug reports \cite{gcc2024, llvm2024}. Delta debugging generates minimal test programs that reproduce the bug, where \( \mathbb{S} \)  represents program variants, \( S \) is a bug-triggering variant, and \( \psi \) verifies the bug's reproducibility.

Delta debugging \cite{zeller1999} assumes search space monotonicity: if a program induces failure \(\psi\), any superset satisfies \(\psi\); if not, no subset satisfies \(\psi\). However, this assumption may not hold in practice.

\textbf{Problem Statement.}  
We improve delta debugging—specifically \texttt{ddmin}-style algorithms \cite{ddmin}—by dynamically evaluating search space monotonicity to eliminate non-failure-inducing test cases.  


\textbf{Prior Work.}
In delta debugging, the foundational \ddmin algorithm~\cite{ddmin} employs a divide-and-conquer approach to minimize failure-inducing programs. Refinements of \ddmin focus on reducing redundancy by minimizing non-failure-inducing tests, enhancing efficiency. Early efforts leveraged outcome caching and strict monotonicity assumptions to reduce redundancy~\cite{zeller2009}. Later methods like HDD~\cite{hdd} and Perses~\cite{perses} aimed to generate only syntactically valid tests but often produced semantically invalid, redundant cases. Recent strategies leverage historical test data to improve effectiveness. For instance, \CHISEL~\cite{chisel} uses reinforcement learning to prioritize test cases likely to pass the property test $\psi$, while ProbDD~\cite{probdd} probabilistically generates test cases based on element success likelihood. However, these approaches may not fully exploit the broader search space, potentially overlooking further opportunities to refine delta debugging.


\textbf{This Work.}
A key challenge in delta debugging is efficiently leveraging historical test data to avoid non-failure-inducing test cases. Our insight is that search space monotonicity enables eliminating such subsets. We propose \emph{Probabilistic Monotonicity Assessment} (\PMA), which integrates with \ddmin-style algorithms to boost efficiency without sacrificing effectiveness. Unlike traditional methods (\CHISEL~\cite{chisel}, \ProbDD~\cite{probdd}), \PMA dynamically models  and assesses search space monotonicity, using a confidence function to probabilistically exclude non-failure-inducing cases and uncover patterns missed by prior work.


\textbf{Evaluation.}
We evaluated \PMA's effectiveness through integration with \CHISEL~\cite{chisel}, comparing against both \CHISEL and \ProbDD~\cite{probdd} (the latter implemented within \CHISEL). Our comprehensive benchmarks demonstrate that \PMA achieves a 59.2\% reduction in processing time, a $3.32\times$ improvement in token deletion rate, and yields 6.7\% smaller final programs on average compared to \CHISEL. When compared to \ProbDD, \PMA reduces processing time by 22.0\%, improves reduction speed by $1.34\times$, and further decreases program size by 3.0\%. These results establish \PMA's superior performance in enhancing delta debugging efficiency while maintaining reduction quality. The findings validate the benefits of probabilistic search space analysis and suggest promising applications in program reduction and bug localization.


\textbf{Contributions.} 
We make the following major contributions:
\begin{itemize}
    \item We introduce \emph{Probabilistic Monotonicity Assessment} (\PMA), a novel approach that enhances delta debugging efficiency through probabilistic modeling of search space monotonicity;
    \item We establish that probabilistic analysis of search space properties (particularly monotonicity) advances delta debugging methodology, opening new research directions in probabilistic program reduction and bug localization; and
    \item We empirically demonstrate \PMA's superiority over state-of-the-art techniques through comprehensive evaluation.
\end{itemize}

\lstset{
    basicstyle=\ttfamily\footnotesize,
    frame=single,
    breaklines=true,
    captionpos=b,
    numbers=left,
    numberstyle=\tiny,
    stepnumber=1,
    numbersep=5pt,
    showstringspaces=false,
    tabsize=4
}

\begin{figure}
\centering
\begin{lstlisting}[language=C]
void main() {
    printf("Line 1\n");
    printf("Line 2\n");
    printf("Line 3\n");
    printf("Line 4\n");
    printf("Line 5\n");
    printf("Line 6\n");
    printf("Line 10\n"); }// a bug where the expected output is "Line 7"
\end{lstlisting}

\vspace*{-2.5ex}
\caption{A motivating example for illustrating the operation of the classic \ddmin algorithm.}
\label{example_program}
\label{fig:example_program}
\Description{This listing shows a simple C program printing numbered lines.}

\vspace*{-4ex}
\end{figure}

\section{Background and Motivation}
\label{sec:back}

We first review delta debugging (\Cref{sec:delta}) and the \ddmin algorithm (\Cref{sec:ddmin}), then demonstrate through an example how eliminating non-failure-inducing tests accelerates \ddmin-style algorithms (\Cref{Sec:Motivating example}).

\newcommand{\included}[1]{\cellcolor[HTML]{00B0F0}s\textsubscript{#1}}
\newcommand{\excluded}[1]{s\textsubscript{#1}}
\newcommand{\testcase}[1]{t\textsubscript{#1}}
\definecolor{mycyan}{HTML}{00B0F0}

\begin{figure*}[t]
\scalebox{1}[0.85]{
\centering
\small
\setlength{\tabcolsep}{3pt}
\begin{tabular}{*{2}{c} | *{2}{c} | *{6}{c} | *{5}{c} | *{2}{c} | *{6}{c} | *{5}{c} | *{2}{c}}

\toprule
\multicolumn{2}{c|}{Test Case}  & \testcase{1} & \testcase{2} & \testcase{3} & \testcase{4} & \testcase{5} & \testcase{6} & \testcase{7} & \testcase{8} & \testcase{9} & \testcase{10} & \testcase{11} & \testcase{12} & \testcase{13} & \testcase{14} & \testcase{15} & \testcase{16} & \testcase{17} & \testcase{18} & \testcase{19} & \testcase{20} & \testcase{21} & \testcase{22} & \testcase{23} & \testcase{24} & \testcase{25} & \testcase{26} & \testcase{27} & \testcase{28} \\
\midrule

\multicolumn{2}{c|}{Granularity ($n$)} & \multicolumn{2}{c|}{ 2} & \multicolumn{6}{c|}{ 4} & \multicolumn{5}{c|}{ 3} & \multicolumn{2}{c|}{2} & \multicolumn{6}{c|}{ 4} & \multicolumn{5}{c|}{ 3} & \multicolumn{2}{c}{2} \\

\midrule
\multirow{8}{*}{\rotatebox[origin=c]{90}{Initial Elements}} 
& \included{1} & \included{1} & \excluded{1} & \included{1} & \excluded{1} & \excluded{1} & \excluded{1} & \excluded{1} & \included{1} & \included{1} & \excluded{1} & \excluded{1} & \excluded{1} & \included{1} & \included{1} & \excluded{1} & \included{1} & \excluded{1} & \excluded{1} & \excluded{1} & \excluded{1} & \included{1} & \included{1} & \excluded{1} & \excluded{1} & \excluded{1} & \included{1} & \included{1} & \excluded{1} \\
& \included{2} & \included{2} & \excluded{2} & \included{2} & \excluded{2} & \excluded{2} & \excluded{2} & \excluded{2} & \included{2} & \included{2} & \excluded{2} & \excluded{2} & \excluded{2} & \included{2} & \included{2} & \excluded{2} & \excluded{2} & \included{2} & \excluded{2} & \excluded{2} & \included{2} & \excluded{2} & \excluded{2} & \excluded{2} & \excluded{2} & \excluded{2} & \excluded{2} & \excluded{2} & \excluded{2} \\
& \included{3} & \included{3} & \excluded{3} & \excluded{3} & \included{3} & \excluded{3} & \excluded{3} & \included{3} & \excluded{3} & \excluded{3} & \excluded{3} & \excluded{3} & \excluded{3} & \excluded{3} & \excluded{3} & \excluded{3} & \excluded{3} & \excluded{3} & \excluded{3} & \excluded{3} & \excluded{3} & \excluded{3} & \excluded{3} & \excluded{3} & \excluded{3} & \excluded{3} & \excluded{3} & \excluded{3} & \excluded{3} \\
& \included{4} & \included{4} & \excluded{4} & \excluded{4} & \included{4} & \excluded{4} & \excluded{4} & \included{4} & \excluded{4} & \excluded{4} & \excluded{4} & \excluded{4} & \excluded{4} & \excluded{4} & \excluded{4} & \excluded{4} & \excluded{4} & \excluded{4} & \excluded{4} & \excluded{4} & \excluded{4} & \excluded{4} & \excluded{4} & \excluded{4} & \excluded{4} & \excluded{4} & \excluded{4} & \excluded{4} & \excluded{4} \\
& \included{5} & \excluded{5} & \included{5} & \excluded{5} & \excluded{5} & \included{5} & \excluded{5} & \included{5} & \included{5} & \excluded{5} & \included{5} & \excluded{5} & \included{5} & \excluded{5} & \excluded{5} & \excluded{5} & \excluded{5} & \excluded{5} & \excluded{5} & \excluded{5} & \excluded{5} & \excluded{5} & \excluded{5} & \excluded{5} & \excluded{5} & \excluded{5} & \excluded{5} & \excluded{5} & \excluded{5} \\
& \included{6} & \excluded{6} & \included{6} & \excluded{6} & \excluded{6} & \included{6} & \excluded{6} & \included{6} & \included{6} & \excluded{6} & \included{6} & \excluded{6} & \included{6} & \excluded{6} & \excluded{6} & \excluded{6} & \excluded{6} & \excluded{6} & \excluded{6} & \excluded{6} & \excluded{6} & \excluded{6} & \excluded{6} & \excluded{6} & \excluded{6} & \excluded{6} & \excluded{6} & \excluded{6} & \excluded{6} \\
& \included{7} & \excluded{7} & \included{7} & \excluded{7} & \excluded{7} & \excluded{7} & \included{7} & \included{7} & \included{7} & \excluded{7} & \excluded{7} & \included{7} & \included{7} & \included{7} & \excluded{7} & \included{7} & \excluded{7} & \excluded{7} & \included{7} & \excluded{7} & \included{7} & \included{7} & \excluded{7} & \included{7} & \excluded{7} & \included{7} & \excluded{7} & \excluded{7} & \excluded{7} \\
& \included{8} & \excluded{8} & \included{8} & \excluded{8} & \excluded{8} & \excluded{8} & \included{8} & \included{8} & \included{8} & \excluded{8} & \excluded{8} & \included{8} & \included{8} & \included{8} & \excluded{8} & \included{8} & \excluded{8} & \excluded{8} & \excluded{8} & \included{8} & \included{8} & \included{8} & \excluded{8} & \excluded{8} & \included{8} & \included{8} & \included{8} & \excluded{8} & \included{8} \\
\midrule
\multicolumn{2}{c|}{$\psi$} & F & F & F & F & F & F & F & T & F* & F* & F* & F* & T & F* & F* & F & F & F & F & F & T & F* & F* & F* & F* & T & F* & F* \\
\bottomrule
\end{tabular}
}
\vspace*{-1.5ex}
\caption{Step-by-step trace of the \ddmin algorithm applied to the motivating example in \Cref{fig:example_program}. Each step includes a test case comprising statements highlighted in \colorbox{mycyan}{cyan}. Test cases are flagged 'T' if they satisfy $\psi$ and 'F' otherwise. Test cases marked with $\text{F}^*$ denote previously attempted (non-failure-inducing) ones.}
\label{tab:ddmin-process}
\Description{It shows the step-by-step outcomes from \ddmin on the running example.}
\label{fig:ddminprocess}
\vspace*{-2.ex}
\end{figure*}

\subsection{Delta Debugging} 
\label{sec:delta}

Delta debugging relies on three core assumptions about the programs being reduced: monotonicity, unambiguity and consistency \cite{zeller1999, ddmin}. We utilize the property test function \( \psi: \mathbb{S} \to \{\true, \false\} \), where \( \mathbb{S} = 2^S \) represents the search space for a program \(S\)—composed of elements or tokens—that satisfies \( \psi(S) = \true \) (\Cref{sec:intro}).

\begin{assumption}[Monotonicity]
\label{def:monotonicity}
A search space $\mathbb{S}$ is \emph{monotone} if, for all $S, S' \in \mathbb{S}$ such that $S \supseteq S'$, then $\psi(S) = \false \implies \psi(S') = \false$, or equivalently, $\psi(S') = \true \implies \psi(S) =  \true$.
\end{assumption}
According to the monotonicity of the search space, any subset of a non-failure-inducing test case will not induce failure, ensuring consistent propagation of the failure property from the test case to all its subsets within the search space \( \mathbb{S} \). 
Conversely, any superset of a failure-inducing test case will also induce the same failure.  

\begin{assumption}[Unambiguity]
\label{def:umabiguity}
A search space $\mathbb{S}$ is \emph{unambiguous} if, for any $S_1, S_2 \in \mathbb{S}$ where both $\psi(S_1) = \true$ and $\psi(S_2) = \true$, it follows that $\psi(S_1 \cap S_2) = \true$.
\end{assumption}
Thus, a failure is caused by a specific subset rather than independently by two disjoint subsets.

\begin{assumption}[Consistency]
\label{def:consistency}
A search space $\mathbb{S}$ is \emph{consistent} if, for any $S \in \mathbb{S}$, $\psi(S)\in\{\true,\false\}$.
\end{assumption}
Despite recent advances in delta debugging aimed at testing only consistent sequences from a given input program \cite{hdd,perses,creduce}, some test cases may still yield inconsistent results due to factors such as syntactic or semantic errors. We examine these effects on \PMA and discuss potential improvements in \Cref{sec:eval}.

In an ideal debugging environment where monotonicity, unambiguity and consistency prevail, if a failure-inducing program \( S \) satisfies \( \psi(S) = \text{true} \), we can always pinpoint the smallest program \( S^* \) in \( \mathbb{S} \) that maintains the failure-inducing property by solving the following optimization problem:
\begin{equation}
S^* = \underset{S \in \mathbb{S}}{\text{argmin}} \left| S \right| \quad \text{s.t.} \quad \psi(S) = \true
\end{equation}
In practice, however, these two assumptions may not hold, making the problem of finding a global optimal solution NP-complete \cite{zellerisolating}. Consequently, the focus of delta debugging shifts toward identifying a locally optimal solution, typically characterized by \emph{1-minimality} \cite{zellerisolating}. A test case \( S^* \) is considered \emph{1-minimal} if removing any single element results in a variant \( S' \) that fails the property test, ensuring that \( S^* \) is the smallest subset that maintains the desired property.

\subsection{The \ddmin Algorithm} 
\label{sec:ddmin}


The \ddmin algorithm \cite{ddmin} isolates the minimal failure-inducing subset of an input program \( S \) through a systematic process that iteratively partitions and tests its subsets. This approach, grounded in Assumptions~\ref{def:monotonicity} and \ref{def:umabiguity}, comprises the following three main steps:
 
\begin{description}
    \item \textbf{Step 1: Initial Partitioning.}
    Initially set \( n = 2 \). Divide \( S \) into \( n \) partitions: \( U = \{u_1, u_2, \dots, u_n\} \).
    Test each partition: for each \( i \) in \( [1, n] \), check if \( \psi(u_i) = \true \).
    If any partition \( u_i \) satisfies \( \psi(u_i) = \true \), update \( S \leftarrow u_i \) and restart at \textbf{Step~1}.

    \item \textbf{Step 2: Complement Testing.}
    If no partition passes the test (\( \forall\ i : \psi(u_i)= \false \)), evaluate the complements: for each \( i \) in \( [1, n] \), test \( \psi(\overline{u_i}) = \psi(S \setminus u_i) \).
    If a complement \( \overline{u_i} \) passes (\( \psi(\overline{u_i}) = \true \)), update \( S \leftarrow \overline{u_i} \), reduce granularity \( n \leftarrow \max(n - 1, 2) \), and restart at \textbf{Step 1}.

    \item \textbf{Step 3: Granularity Adjustment.}
    If all tests in \textbf{Steps 1} and \textbf{2} fail and \( n < |S| \), increase granularity: \( n \leftarrow \min(2n, |S|) \) and return to \textbf{Step 1}.
    If \( n \geq |S| \), terminate the process and return \( S \) as the minimal failure-inducing subset.
\end{description}

\subsection{A Motivating Example}
\label{Sec:Motivating example}

To demonstrate \ddmin's operation and reveal redundant, non-failure-inducing test cases, we analyze its application to a C program (Figure~\ref{example_program}). This program, intended to sequentially print numbers from "1" to "7," erroneously outputs "10" instead of "7" due to a coding error. \ddmin identifies the minimal error-reproducing subset $\{s_1, s_8\}$, where $s_1$ prevents compilation errors and $s_8$ produces the faulty output. The program's eight statements are labeled $s_i$ (line $i$).

\Cref{fig:ddminprocess} illustrates the \ddmin algorithm on our example program, detailing the test sequence for each case \(t_i\) at every step. To ultimately isolate the minimal failure-inducing sequence \(t_{28}=\{s_1, s_8\}\), \ddmin initially splits \( S = \{s_1, s_2, ... , s_8\} \) into two partitions ($t_1$-$t_2$). 
It then increases the partitions to four (\(n=4\)) when this first division 
fails to isolate the bug. 
This increase in granularity ($t_3$-$t_6$) still does not reveal a smaller failure-inducing test case, leading to complement testing in Step 8 that identifies a significant sequence \( t_8 = \{s_1, s_2, s_5, s_6, s_7, s_8\} \) reproducing the bug. 
Steps 9-12 further divide \(t_8\) into three singleton subsets, all proving unsuccessful earlier (indicated by $\text{F}^*)$. 
A critical breakthrough occurs at Step 13, where testing the complement of \( \{s_5, s_6\} \) relative to \(t_8\) further refines the sequence to \( t_{13} = \{s_1, s_2, s_7, s_8\} \). Additional tests narrow it down to \(t_{21} = \{s_1, s_7, s_8\}\) in Step 21, and ultimately to \(t_{26} = \{s_1, s_8\}\) in Step 26, isolating the minimal failure-inducing sequence. The process concludes after Steps 27-28, as \( n \geq |S| \) when $n=|S|=2$, effectively reducing the original program from eight statements to just two.

This example highlights a significant limitation of \ddmin \cite{ddmin} and its derivatives, such as \CHISEL \cite{chisel}, in practical debugging scenarios. \ddmin-style algorithms often produce numerous non-failure-inducing test cases, including those previously attempted (marked by $\text{F}^*$ in \Cref{fig:ddminprocess}), leading to suboptimal efficiency in program reduction, particularly in large-scale applications. This inefficiency underscores the necessity for a more streamlined approach to eliminating redundant non-failure-inducing tests.

\section{Probabilistic Monotonicity Assessment}

To improve \ddmin-style algorithms by eliminating redundant non-failure-inducing tests, we present a novel Probabilistic Monotonicity Assessment (\PMA) approach. This approach utilizes a probabilistic model to quantify the monotonicity of the search space, dynamically assessing monotonicity based on historical test data to strategically remove tests unlikely to reproduce the failure when debugging a given input program. 
Initially, we introduce this probabilistic model in \Cref{sec:model}, provide a theoretical justification for our approach in \Cref{sec:just}, and finally, revisit our motivating example in \Cref{sec:revisit-ex} to demonstrate the effectiveness of our new approach.

\begin{figure}[th]
\centering

\begin{mdframed}
    \scalebox{0.85}{
	\begin{minipage}{\textwidth}
   \begin{tabular}{l}
 \begin{tabular}{l}
        \colorbox{orange!50}{\textbf{Model Update}} \textbf{(After Each  Test Case $T$ is Executed):} \\
         \vspace*{-2.5ex}
 \\
  \hspace{3ex} (A1) Assess $\monoObserver(T)$\\
        \hspace{3ex} (A2) Update    $\monoCount \leftarrow \monoUpdate(T)$\\
       \hspace{3ex} (A3) Recalibrate   Confidence Function $\confun(\monoCount)$
        \\
        \end{tabular}
        \vspace{1.5ex}
         \\
 \begin{tabular}{@{\hspace{5ex}}ll}
	\multicolumn{2}{l}{\colorbox{orange!50}{\textbf{Model Application}} \textbf{(Before Each Test Case $T$ is Executed):}}  \\
 \vspace*{-2.5ex}
 \\
   (B1) & Evaluate \monoCheck($T$)    \\
     (B2)  & Evaluate $\skipTest(\monoCount) = \begin{cases}
\true &   \confun(\monoCount) > \mathcal{U}(0, 1)    \\
\false & \text{otherwise}  
\end{cases}$ \\
      (B3) & \textbf{if} $\monoCheck(T) \wedge  \skipTest(\monoCount)  =\true$  \textbf{then}\\
      (B4) & \hspace{3ex} Skip $T$ \\
(B5) & \textbf{else} \\
(B6) & \hspace{3ex} Execute $T$
\end{tabular}
      \end{tabular}
	\end{minipage}
    }
\end{mdframed}

\vspace*{-1ex}
\caption{The probabilistic model for eliminating redundant non-failure-inducing tests.
\label{fig:model}
}
\vspace*{-2ex}
\end{figure}

\subsection{The Probabilistic Model} 
\label{sec:model}

\PMA is designed to enhance a \ddmin-style algorithm by eliminating redundant tests using a simple yet effective probabilistic model, as shown in \Cref{fig:model}. While the \ddmin-style algorithm operates normally, it integrates with \(\text{PMA}\), which determines whether to skip or execute each newly generated test case. \(\text{PMA}\) intervenes both after a test case \(T\) has been executed, to update the model, and just before its execution, to apply the model, as detailed below:
\begin{itemize}
    \item \colorbox{orange!50}{Model Update (\Cref{sec:modelupdate}).}
    After each test  \( T \) is executed, \PMA updates the search space's monotonicity by first assessing \(\monoObserver(T)\) (line A1). It then updates \(\monoCount\) using \(\monoCount \leftarrow \monoUpdate(T)\), which quantifies the net difference between observed monotonicity compliances and violations from previous tests (line A2). This updated measure refines the confidence function, \(\confun(\monoCount)\), evaluating current monotonicity  (line A3).
  
    \begin{sloppypar} 
    \item \colorbox{orange!50}{Model Application (\Cref{sec:modelapp}).}
    Before executing each test \( T \), \PMA evaluates the search space's monotonicity through \(\monoCheck(T)\) outcomes (line B1) to determine if \( T \) can be classified as redundant. 
    Additionally, \(\skipTest(\monoCount)\) is assessed using the confidence function \(\confun(\monoCount)\) to probabilistically decide if \( T \) should be skipped (line B2). \PMA will only skip \( T \) if both \(\monoCheck(T)\) and \(\skipTest(\monoCount)\) concur on this decision (lines B3-B6).
    \end{sloppypar}

\end{itemize}

Note that the input program $S$, to be reduced, is initially treated as a test case with $\psi(S) = \true$.

\subsubsection{Model Update}
\label{sec:modelupdate}

The essence of our probabilistic model lies in capturing the search space's monotonicity in a \ddmin-style algorithm. Traditionally viewed as a binary property in formal logic \cite{programanalysis}—where a search space is either monotone or not—our approach introduces a probabilistic interpretation to handle real-world software system complexities more effectively.

In practice, software systems often display non-deterministic behavior due to factors like concurrency, environmental variables, or hidden dependencies. Furthermore, test results may be affected by noise or intermittent failures, obscuring the true nature of the search space. Consequently, a strictly binary view of monotonicity can result in premature or overly conservative conclusions.

To address these challenges, we define monotonicity \(M\) in probabilistic terms:
\begin{equation}
P(M \mid \mathcal{E}) = \confun(\monoCount)
\end{equation}
where \(P(M \mid \mathcal{E})\) represents the probability of monotonicity given the current evidence \( \mathcal{E} \) (captured by historical test data). This probability is estimated using our confidence function \( \confun(\monoCount) \), with \(\monoCount\) quantifying the net difference between monotonicity compliances and violations observed.

When assessing monotonicity, we avoid setting \(P(M \mid \mathcal{E})\) to binary extremes like one or zero. Instead, by adjusting \(\monoCount\) and \(\confun(\monoCount)\), we gradually modulate the probability. This continuous approach integrates new evidence smoothly and aligns with the formal definition of monotonicity (Assumption~\ref{def:monotonicity}), offering a practical way to handle uncertainties in software debugging.

Below, we detail how \(\text{PMA}\) updates its probabilistic model after executing each test case \(T\) generated by the \ddmin-style algorithm. This includes assessing \(\monoObserver(T)\),  
updating ``\(\monoCount \leftarrow \monoUpdate(T)\)'' and 
recalibrating 
\(\confun(\monoCount)\), as illustrated in \Cref{fig:model} (lines A1-A3).

\paragraph{\bf Assessing $\monoObserver(T)$.}
\label{sec:monoObserver}

We focus on assessing the monotonicity of the search space by concentrating on failure-inducing test cases, which are fewer than non-failure-inducing ones, thus minimizing computational overhead. We define \(\supTest(S)\) as the set of test cases that are supersets of \(T\) and were executed before \(T\). The evaluation of \(\monoObserver(T)\) proceeds as follows:

\begin{equation}
\hspace{-0.75em}
\label{eq:mu}
\monoObserver(T) = 
\begin{cases}
\begin{cases}
\false & \exists\ T'\in \supTest(T) : \\
       & \quad \psi(T') = \false  \\
\true  & \text{otherwise} 
\end{cases} & \psi(T) = \true \\
\top & \text{otherwise}
\end{cases}
\end{equation}

Under Assumption~\ref{def:monotonicity}, a new failure-inducing test case \(T\), where \(\psi(T) = \true\), preserves the search space's monotonicity if \\ \(\monoObserver(T) = \true\) and violates it if \(\monoObserver(T) = \false\).

\paragraph{\bf Computing $\monoCount \leftarrow \monoUpdate(T)$.}
\label{sec:monoUpdate}

\(\monoCount\), which quantifies the net difference between monotonicity compliances and violations, is initialized to zero when the \(\ddmin\)-style algorithm begins. After each failure-inducing test case, we use \(\monoObserver(S)\) to determine if the test outcome conforms to the expected monotonicity, and we update \(\monoCount\) accordingly:
\begin{equation}
\monoCount \leftarrow \monoUpdate(T) = 
\begin{cases}
\monoCount + 1 &   \monoObserver(T) = \true \\
\monoCount - 1 &   \monoObserver(T) = \false \\
\monoCount  &  \monoObserver(T) = \top
\end{cases}
\end{equation}
By the definition of $ \monoObserver(T)$,
\( \monoCount \) remains unchanged if \( T \) is a non-failure-inducing test case.

Under Assumption~\ref{def:monotonicity}, if a test case does not induce a failure, its non-failure-inducing property should consistently propagate to all its subsets.  When \( T \) emerges as a new failure-inducing test case, we evaluate monotonicity compliance: if \( \monoObserver(T) = \true \), signaling compliance, we increase \( \monoCount \) by one. If \( \monoObserver(T) = \false \), indicating a violation, we decrease \( \monoCount \) by one.

\paragraph{\bf Recalibrating  $\confun(\monoCount)$.}
\label{confidence function}
\label{sec:confun}

We use this confidence function to gauge the search space's monotonicity:
\begin{equation}
\label{eq:confidence_function}
\confun(\monoCount) = \frac{1}{1 + e^{-\monoCount}}
\end{equation}
When \PMA starts, $\monoCount=0$, setting $\confun(\monoCount)=0.5$, indicating neutral confidence in the search space's monotonicity. 
As each new test case is executed, \(\confun(\monoCount)\) is recalibrated based on the updated \(\monoCount\), 
adapting the confidence level to observed monotonicity.

Our confidence function, \( \confun(\monoCount) \), is grounded in the Logistic Growth Model~\cite{logisitc}, which we have adapted to quantify our belief in the search space's monotonicity during the debugging process. This function offers several desirable properties:
\begin{enumerate}
    \item \textbf{Bounded Output:} \( \confun(\monoCount) \in (0,1) \).
    \item \textbf{Monotonicity:} \( \frac{d\confun}{d\monoCount} > 0 \).
    \item \textbf{Symmetry:} \( \confun(x) = 1 - \confun(-x) \).
    \item \textbf{Convergence:} \( \lim_{\monoCount \to \infty} \confun(\monoCount) = 1 \) and \( \lim_{\monoCount \to -\infty} \confun(\monoCount) \newline= 0 \), illustrating the asymptotic behavior of 
    \( \confun(\monoCount) \).

\end{enumerate}

\subsubsection{Model Application}
\label{sec:modelapp}

The confidence function \( \confun(\monoCount) \) is central to our debugging approach, quantifying the search space's monotonicity and enabling \PMA efficiently to skip redundant tests.

\begin{sloppypar}
Below, we detail how \(\text{PMA}\) applies its probabilistic model just before executing each test case \(T\) generated by the \ddmin-style algorithm. This involves evaluating \(\monoCheck(T)\) and  \(\skipTest(\monoCount)\) to decide if the test should be skipped, as depicted in \Cref{fig:model} (lines B1-B6).
\end{sloppypar}

\paragraph{\bf Evaluating $\monoCheck(T)$.}

To decide if test case \(T\) can be skipped, we assess whether its execution could violate monotonicity. We continue to write \(\supTest(S)\) to represent the set of test cases that are supersets of \(T\) and were previously executed. The criterion for \(\monoCheck(T)\) is:
\begin{equation}
\hspace{-0.944em}
\label{eq:monocheck}
\monoCheck(T) = 
\begin{cases}
\true &   \exists\ T' \in \supTest(T) : \psi(T') = \false \\
\false & \text{otherwise}
\end{cases}
\end{equation}
If \(\monoCheck(T)\) returns true, it indicates that a previously executed superset of \(T\) was non-failure-inducing. By Assumption~\ref{def:monotonicity}, \(T\) can then be eliminated as redundant. Otherwise, \(T\) can proceed without violating monotonicity.

\begin{sloppypar}
Note that some regenerated non-failure-inducing test cases, marked by $\texttt{F}^*$ in \Cref{fig:ddminprocess}, can lead to unnecessary overhead if avoided entirely. However, since each test case inherently includes itself, such redundancies are probabilistically eliminated. \PMA's effectiveness in reducing redundant tests
including these as a special case is demonstrated in our evaluation (\Cref{sec:RQ1}).
\end{sloppypar}

\paragraph{\bf Evaluating $\skipTest(\monoCount)$.}

If \(\monoCheck(T) = \true\), there is a potential to skip \(T\) as a redundant, non-failure-inducing test. The final decision depends on evaluating \(\skipTest(\monoCount)\), as outlined in \Cref{fig:model} (line B2) and reproduced below, guided by our confidence function \(\confun(\monoCount)\) defined as follows:

\begin{equation}
\skipTest(\monoCount) = \begin{cases}
\true &   \confun(\monoCount) > \mathcal{U}(0, 1)    \\
\false & \text{otherwise}  
\label{eq:skiptest}
\end{cases}   
\end{equation}
The test case \(T\) will be skipped if \(\skipTest(\monoCount)\) agrees (positively) with \(\monoCheck(T)\) as outlined in lines B3-B6 of \Cref{fig:model}; otherwise, it will be executed. Here, \( \mathcal{U}(0, 1) \) represents a random draw from a uniform distribution over the interval (0, 1), ensuring impartial decisions.
The model uses \( \confun(\monoCount) \) as a dynamic threshold integrating historical test data. A high \( \confun(\monoCount)\) value enables strategic omission of certain subsets of a non-failure-inducing test case, while a low value necessitates comprehensive testing.


\subsubsection{Discussion}

Leveraging feedback from the confidence function \( \confun(\monoCount) \) about search space monotonicity, \PMA optimizes a \ddmin-style algorithm with an adaptive testing strategy:
\begin{enumerate}
    \item \textbf{Iterative Updates:} Each failure-inducing test updates \( \monoCount \), continuously adjusting \( \confun(\monoCount) \) to refine the testing strategy based on evolving insights into the search space.  

    \item \textbf{Selective Testing:} Tests are chosen based on their potential impact on the efficiency and effectiveness of
    the \ddmin-style algorithm and the prevailing confidence level, prioritizing those likely to provide significant insights or addressing the most significant uncertainties.
\end{enumerate}

By leveraging \( \confun(\monoCount) \),our approach boosts the debugging process's efficiency and focuses efforts where they are most needed, guided by a probabilistic understanding of the search space's monotonicity. 
This strategy minimizes unnecessary tests while ensuring effective identification of minimal failure-inducing test cases.

\subsection{Theoretical Justification}
\label{sec:just}

We provide a brief theoretical justification for our \PMA approach, which can be seen as a special case of two established theoretical models discussed below.

\subsubsection{Reliability of Confidence Estimation}

The \emph{Law of Large Numbers (LLN)}~\cite{lln} ensures sample averages converge to the expected value, providing a theoretical foundation for analyzing random variables.

\begin{theorem}[LLN~\cite{lln}]
Let $X_1, X_2, ..., X_n$ be independent and identically distributed Bernoulli random variables with an expected value $\mathbb{E}[X_i] = \mu$. Then for any $\epsilon > 0$, we have:
\[\lim_{n \to \infty} P\left(\left|\frac{1}{n}\sum_{i=1}^n X_i - \mu\right| < \epsilon\right) = 1\]
\end{theorem}


Applying \(\monoUpdate(T)\) updates \(\monoCount\) when executing the \(i\)-th failure-inducing test case (\(\psi(T) = \true\)), where the event \(X_i\) is 1 for compliance and 0 for violation. Here, \(\mu\) denotes the probability of monotonicity in the search space. As \( n \) increases, the empirical average \( \frac{1}{n} \sum_{i=1}^{n} X_i \) converges to \(\mu\), ensuring that \PMA's estimates increasingly reflect the true probability of monotonicity. As \( n \to \infty \), with \(\monoCount \approx n\mu\), the confidence function \(\confun(\monoCount)\) translates accumulated monotonicity evidence into a probability measure.



\subsubsection{Effectiveness of Redundancy Elimination}

\PMA's effectiveness in skipping redundant tests at high $\confun(\monoCount)$ values is underpinned by information gain, leveraging Shannon entropy—a metric that measures uncertainty in information systems~\cite{shannon1948mathematical}.

\begin{theorem}[Information Gain~\cite{shannon1948mathematical}]
Information gain\linebreak $IG(X; Y)$ from observing a random variable $Y$ about another variable $X$ is defined as follows:
\[
IG(X; Y) = H(X) - H(X|Y)
\]
where \( H \) represents Shannon entropy, which quantifies the level of uncertainty or expected information content in \( X \) both when considered independently and in the context of knowing the value of \( Y \).
\end{theorem}
Here, $IG(X; Y)$ measure quantifies how observing \( Y \) reduces uncertainty or boosts knowledge about \( X \), thus enhancing our understanding relative to not knowing \( Y \).

In our \PMA approach, \(\monoCount\) represents the net difference between monotonicity compliance and violation. Here, $X$ denotes the inherent monotonicity of the search space, and $Y$ represents the outcome of an executed test case, classified as failure-inducing or not by \(\psi\). 
As $|\monoCount|$ increases, reflecting more executed tests and enhanced confidence through $\confun(\monoCount)$, $H(X)$ decreases due to reduced uncertainty about monotonicity. However, $H(X|Y)$—the entropy of $X$ given a new test outcome $Y$—remains close to $H(X)$, suggesting that additional tests contribute little new information. Consequently, when $|\monoCount|$ is large, the following relationship holds:
\[IG(X;Y) = H(X) - H(X|Y) \approx 0\]
This minimal information gain justifies skipping redundant tests at high $\confun(\monoCount)$ levels, as further testing offers minimal additional insight in identifying the minimal failure-inducing test.

\subsection{Revisiting the Motivating Example}
\label{sec:revisit-ex}

\newcommand{\NA}{\text{---}}
\newcommand{\Execute}{\cellcolor{red!20}\textcolor{red!70!black}{F}}
\newcommand{\Skip}{\cellcolor{green!20}\textcolor{green!70!black}{T}}

\newcommand{\confunvalue}{0}
\newcommand{\setconfunvalue}[1]{\renewcommand{\confunvalue}{#1}}

\newcommand{\coloredCell}[2]{%
  \cellcolor{#2!20}\textcolor{#2!70!black}{#1}%
}

\begin{figure*}[t]
\hspace*{-.3ex}\scalebox{0.85}[0.6]{
\centering
\small
\setlength{\tabcolsep}{3pt}
\begin{tabular}{*{2}{c} | *{2}{c} | *{6}{c} | *{5}{c} | *{2}{c} | *{6}{c} | *{5}{c} | *{2}{c}}

\toprule
\multicolumn{2}{c|}{Test Case}  & \testcase{1} & \testcase{2} & \testcase{3} & \testcase{4} & \testcase{5} & \testcase{6} & \testcase{7} & \testcase{8} & \testcase{9} & \testcase{10} & \testcase{11} & \testcase{12} & \testcase{13} & \testcase{14} & \testcase{15} & \testcase{16} & \testcase{17} & \testcase{18} & \testcase{19} & \testcase{20} & \testcase{21} & \testcase{22} & \testcase{23} & \testcase{24} & \testcase{25} & \testcase{26} & \testcase{27} & \testcase{28} \\
\midrule

\multicolumn{2}{c|}{Granularity ($n$)} & \multicolumn{2}{c|}{ 2} & \multicolumn{6}{c|}{ 4} & \multicolumn{5}{c|}{ 3} & \multicolumn{2}{c|}{ 2} & \multicolumn{6}{c|}{ 4} & \multicolumn{5}{c|}{ 3} & \multicolumn{2}{c}{ 2} \\

\midrule
\multirow{8}{*}{\rotatebox[origin=c]{90}{Initial Elements}} 
& \included{1} & \included{1} & \excluded{1} & \included{1} & \excluded{1} & \excluded{1} & \excluded{1} & \excluded{1} & \included{1} & \included{1} & \excluded{1} & \excluded{1} & \excluded{1} & \included{1} & \included{1} & \excluded{1} & \included{1} & \excluded{1} & \excluded{1} & \excluded{1} & \excluded{1} & \included{1} & \included{1} & \excluded{1} & \excluded{1} & \excluded{1} & \included{1} & \included{1} & \excluded{1} \\
& \included{2} & \included{2} & \excluded{2} & \included{2} & \excluded{2} & \excluded{2} & \excluded{2} & \excluded{2} & \included{2} & \included{2} & \excluded{2} & \excluded{2} & \excluded{2} & \included{2} & \included{2} & \excluded{2} & \excluded{2} & \included{2} & \excluded{2} & \excluded{2} & \included{2} & \excluded{2} & \excluded{2} & \excluded{2} & \excluded{2} & \excluded{2} & \excluded{2} & \excluded{2} & \excluded{2} \\
& \included{3} & \included{3} & \excluded{3} & \excluded{3} & \included{3} & \excluded{3} & \excluded{3} & \included{3} & \excluded{3} & \excluded{3} & \excluded{3} & \excluded{3} & \excluded{3} & \excluded{3} & \excluded{3} & \excluded{3} & \excluded{3} & \excluded{3} & \excluded{3} & \excluded{3} & \excluded{3} & \excluded{3} & \excluded{3} & \excluded{3} & \excluded{3} & \excluded{3} & \excluded{3} & \excluded{3} & \excluded{3} \\
& \included{4} & \included{4} & \excluded{4} & \excluded{4} & \included{4} & \excluded{4} & \excluded{4} & \included{4} & \excluded{4} & \excluded{4} & \excluded{4} & \excluded{4} & \excluded{4} & \excluded{4} & \excluded{4} & \excluded{4} & \excluded{4} & \excluded{4} & \excluded{4} & \excluded{4} & \excluded{4} & \excluded{4} & \excluded{4} & \excluded{4} & \excluded{4} & \excluded{4} & \excluded{4} & \excluded{4} & \excluded{4} \\
& \included{5} & \excluded{5} & \included{5} & \excluded{5} & \excluded{5} & \included{5} & \excluded{5} & \included{5} & \included{5} & \excluded{5} & \included{5} & \excluded{5} & \included{5} & \excluded{5} & \excluded{5} & \excluded{5} & \excluded{5} & \excluded{5} & \excluded{5} & \excluded{5} & \excluded{5} & \excluded{5} & \excluded{5} & \excluded{5} & \excluded{5} & \excluded{5} & \excluded{5} & \excluded{5} & \excluded{5} \\
& \included{6} & \excluded{6} & \included{6} & \excluded{6} & \excluded{6} & \included{6} & \excluded{6} & \included{6} & \included{6} & \excluded{6} & \included{6} & \excluded{6} & \included{6} & \excluded{6} & \excluded{6} & \excluded{6} & \excluded{6} & \excluded{6} & \excluded{6} & \excluded{6} & \excluded{6} & \excluded{6} & \excluded{6} & \excluded{6} & \excluded{6} & \excluded{6} & \excluded{6} & \excluded{6} & \excluded{6} \\
& \included{7} & \excluded{7} & \included{7} & \excluded{7} & \excluded{7} & \excluded{7} & \included{7} & \included{7} & \included{7} & \excluded{7} & \excluded{7} & \included{7} & \included{7} & \included{7} & \excluded{7} & \included{7} & \excluded{7} & \excluded{7} & \included{7} & \excluded{7} & \included{7} & \included{7} & \excluded{7} & \included{7} & \excluded{7} & \included{7} & \excluded{7} & \excluded{7} & \excluded{7} \\
& \included{8} & \excluded{8} & \included{8} & \excluded{8} & \excluded{8} & \excluded{8} & \included{8} & \included{8} & \included{8} & \excluded{8} & \excluded{8} & \included{8} & \included{8} & \included{8} & \excluded{8} & \included{8} & \excluded{8} & \excluded{8} & \excluded{8} & \included{8} & \included{8} & \included{8} & \excluded{8} & \excluded{8} & \included{8} & \included{8} & \included{8} & \excluded{8} & \included{8} \\

\midrule

\multicolumn{2}{c|}{$\psi$} & \multicolumn{3}{c|}{F} & \multicolumn{2}{c|}{\NA} & \multicolumn{2}{c|}{F} & \multicolumn{1}{c|}{T} & \multicolumn{1}{c|}{F} & \multicolumn{3}{c|}{\NA} & \multicolumn{1}{c|}{T} & \multicolumn{2}{c|}{\NA} & \multicolumn{1}{c|}{F} & \multicolumn{3}{c|}{\NA} & \multicolumn{1}{c|}{F} & \multicolumn{1}{c|}{T} & \multicolumn{4}{c|}{\NA} & \multicolumn{1}{c|}{T} & \multicolumn{2}{c}{\NA} \\

\midrule

\multicolumn{2}{c|}{$\monoObserver(t_i)$}  & \multicolumn{7}{c|}{\NA} & \multicolumn{1}{c|}{\coloredCell{T}{green}} & \multicolumn{4}{c|}{\NA} & \multicolumn{1}{c|}{\coloredCell{T}{green}} & \multicolumn{7}{c|}{\NA} & \multicolumn{1}{c|}{\coloredCell{T}{green}} & \multicolumn{4}{c|}{\NA} & \multicolumn{1}{c|}{\coloredCell{T}{green}} & \multicolumn{2}{c}{\NA}\\

\midrule

\multicolumn{2}{c|}{$\monoCount=\monoUpdate(t_i)$}     & \multicolumn{8}{c|}{0} & \multicolumn{5}{c|}{1} & \multicolumn{8}{c|}{2} & \multicolumn{5}{c|}{3} & \multicolumn{2}{c}{4} \\

\midrule

\multicolumn{2}{c|}{$\confun(\monoCount)$} & \multicolumn{8}{c|}{0.5} & \multicolumn{5}{c|}{0.73} & \multicolumn{8}{c|}{0.88} & \multicolumn{5}{c|}{0.95} & \multicolumn{2}{c}{0.98} \\

\midrule

\multicolumn{2}{c|}{$\mathcal{U}(0, 1)$} & 
\multicolumn{2}{c|}{\NA} & 
\multicolumn{1}{c|}{\coloredCell{0.66}{green}} & 
\multicolumn{1}{c|}{\coloredCell{0.41}{red}} & 
\multicolumn{1}{c|}{\coloredCell{0.17}{red}} & 
\multicolumn{1}{c|}{\coloredCell{0.81}{green}} & 
\multicolumn{2}{c|}{\NA} & 
\multicolumn{1}{c|}{\coloredCell{0.85}{green}} & 
\multicolumn{1}{c|}{\coloredCell{0.66}{red}} & 
\multicolumn{1}{c|}{\coloredCell{0.21}{red}} & 
\multicolumn{1}{c|}{\coloredCell{0.49}{red}} & 
\NA & 
\multicolumn{1}{c|}{\coloredCell{0.81}{red}} & 
\multicolumn{1}{c|}{\coloredCell{0.72}{red}} & 
\multicolumn{1}{c|}{\coloredCell{0.93}{green}} & 
\multicolumn{1}{c|}{\coloredCell{0.73}{red}} & 
\multicolumn{1}{c|}{\coloredCell{0.23}{red}} & 
\multicolumn{1}{c|}{\coloredCell{0.44}{red}} & 
\multicolumn{2}{c|}{\NA} & 
\multicolumn{1}{c|}{\coloredCell{0.67}{red}} & 
\multicolumn{1}{c|}{\coloredCell{0.21}{red}} & 
\multicolumn{1}{c|}{\coloredCell{0.24}{red}} & 
\multicolumn{1}{c|}{\coloredCell{0.92}{red}} & 
\NA & 
\multicolumn{1}{c|}{\coloredCell{0.79}{red}} & 
\multicolumn{1}{c}{\coloredCell{0.97}{red}} \\

\midrule

\multicolumn{2}{c|}{$\monoCheck(t_i)$} & \multicolumn{2}{c|}{F} & \multicolumn{4}{c|}{T} & \multicolumn{2}{c|}{F} & \multicolumn{4}{c|}{T} & \multicolumn{1}{c|}{F} & \multicolumn{6}{c|}{T} & \multicolumn{2}{c|}{F} & \multicolumn{4}{c|}{T} & \multicolumn{1}{c|}{F} & \multicolumn{2}{c}{T} \\

\midrule

\multicolumn{2}{c|}{$\skipTest(\monoCount)$} & \multicolumn{3}{c|}{\Execute} & \multicolumn{2}{c|}{\Skip} & \multicolumn{4}{c|}{\Execute} & \multicolumn{3}{c|}{\Skip} & \multicolumn{1}{c|}{\Execute} & \multicolumn{2}{c|}{\Skip} & \multicolumn{1}{c|}{\Execute} & \multicolumn{3}{c|}{\Skip} & \multicolumn{2}{c|}{\Execute} & \multicolumn{4}{c|}{\Skip} & \multicolumn{1}{c|}{\Execute} & \multicolumn{2}{c}{\Skip} \\

\midrule

\multicolumn{2}{c|}{Skipped Test (\xmark)} & \multicolumn{3}{c|}{} & \multicolumn{1}{c|}{\xmark} & \multicolumn{1}{c|}{\xmark} & \multicolumn{4}{c|}{} & \multicolumn{1}{c|}{\xmark} & \multicolumn{1}{c|}{\xmark} & \multicolumn{1}{c|}{\xmark} & \multicolumn{1}{c|}{} & \multicolumn{1}{c|}{\xmark} & \multicolumn{1}{c|}{\xmark} & \multicolumn{1}{c|}{} & \multicolumn{1}{c|}{\xmark} & \multicolumn{1}{c|}{\xmark} & \multicolumn{1}{c|}{\xmark} & \multicolumn{2}{c|}{} & \multicolumn{1}{c|}{\xmark} & \multicolumn{1}{c|}{\xmark} & \multicolumn{1}{c|}{\xmark} & \multicolumn{1}{c|}{\xmark} & \multicolumn{1}{c|}{} & \multicolumn{1}{c|}{\xmark} & \multicolumn{1}{c|}{\xmark}\\

\bottomrule
\end{tabular}
}
\vspace*{-1.5ex}
\caption{A possible step-by-step trace of \PMA integrated with \ddmin, applied to the motivating example in \Cref{fig:example_program}. The top part replicates the \ddmin trace from \Cref{fig:ddminprocess}, while the bottom part displays the values of functions used by \PMA, with skipped redundant tests indicated in the last row.
}
\label{tab:PMA-process}
\Description{It shows the step-by-step outcomes from \PMA on the running example.}
\label{fig:PMAprocess}
\vspace*{-2ex}
\end{figure*}

Using our motivating example (\Cref{fig:example_program}), we demonstrate \PMA's operation. While \Cref{fig:ddminprocess} showed \ddmin's redundant tests, \Cref{fig:PMAprocess} illustrates \PMA's improvements by eliminating them. The figure presents a step-by-step trace using confidence function $\confun(\monoCount)$ (\Cref{eq:confidence_function}), comparing \ddmin's approach (top) with \PMA's optimizations (bottom), where \xmark's mark skipped redundant tests.

Initially, \(\monoCount = 0\), meaning no prior knowledge or historical test data is assumed. At this point, \(\confun(\monoCount)\) evaluates to \(\confun(0) = \frac{1}{1 + e^{0}} = 0.5\), indicating neutral confidence, with an equal probability of predicting or dismissing the monotonicity of the search space.

\begin{sloppypar}
Initially, \ddmin processes $S = \{s_1, ..., s_8\}$ with $\psi(S) = \true$. Since $\supTest(t_1) = \supTest(t_2) = \{S\}$ yields $\monoCheck(t_1) = \monoCheck(t_2) = \false$, \PMA directs \ddmin to execute the two partitions, $t_1$ and $t_2$. Neither reproduces the bug ($\psi(t_1) = \psi(t_2) = \false$), so no model update occurs.
\end{sloppypar}

Let us see how \PMA analyzes test case \( t_3 \) to determine whether to skip or execute it. Given \( \supTest(t_3) = \{S, t_1\} \) and \( \psi(t_1) = \false \), we obtain \( \monoCheck(t_3) = \true \), suggesting \( t_3 \) might be skipped based on monotonicity. However, skipping is probabilistic rather than deterministic. Since \( \confun(0) = 0.5 \) and \( \mathcal{U}(0,1) = 0.66 \), \(\skipTest(t_3)\) returns \false, leading to the execution of \( t_3 \).

For test cases $t_4$–$t_6$, \PMA skips $t_4$ and $t_5$ but executes $t_6$. Since $t_6$ does not reproduce the bug ($\psi(t_6) = \false$), our model remains unchanged. For $t_7$ and $t_8$, $\supTest(t_7) = \supTest(t_8) = \{S\}$, yielding $\monoCheck(t_7) = \monoCheck(t_8) = \false$ and necessitating their execution. Notably, $t_8$ emerges as a smaller test case that reproduces the bug. This discovery prompts \PMA to update its probabilistic model, following lines A1–A3 of Figure~\ref{fig:model} and described in \Cref{sec:modelupdate}. With $\supTest(t_8) = \{S\}$, we have $\monoObserver(\true) = T$, updating $\monoCount$ to 1 and recalculating $\confun(1) = 0.73$.

For the next five test cases, $t_9$ to $t_{13}$, \PMA is guided by its model to execute $t_9$ and $t_{13}$ while skipping $t_{10}$ to $t_{12}$. Of the two executed test cases, only $t_{13}$ reproduces the bug, prompting another model update: $\monoCount$ is incremented to 2, leading to a recalculated $\confun(2) = 0.88$.

During the rest of this debugging process, \PMA eventually directs \ddmin to execute only four more test cases: $t_{16}$, $t_{20}$, $t_{21}$, and $t_{26}$, with the latter two reproducing the bug. The model continues to update $\monoCount$ and $\confun(\monoCount)$ based on the outcomes of $t_{21}$ and $t_{26}$. Ultimately, \PMA, working in conjunction with \ddmin, successfully identifies \{s1, s8\} as the minimal failure-inducing test case.

In this simple illustrative example, \PMA detected no monotonicity violations, though our model readily adapts when they occur. Compared to \ddmin, \PMA skipped 16 of 28 tests (57\%) by adaptively focusing on bug-revealing cases via the confidence function, avoiding unnecessary tests.

\section{Evaluation}
\label{sec:eval}

Our evaluation assesses the efficiency of \PMA compared to two leading approaches, \CHISEL \cite{chisel} and \ProbDD \cite{probdd}, specifically focusing on C programs. We will also discuss the limitations of our approach and provide insights into potential improvements. The following two research questions guide our study:

\begin{itemize}
\item \textbf{RQ1:} How does \PMA improve program reduction efficiency compared to \CHISEL?
\item \textbf{RQ2:} How does \PMA improve program reduction efficiency compared to \ProbDD?
\end{itemize}

\subsection{Methodology}

We demonstrate the superiority of \PMA over two leading baselines: \CHISEL and \ProbDD. \CHISEL \cite{chisel}, a well-established program reduction tool for C programs, employs \ddmin to generate test cases and uses reinforcement learning to prioritize those likely to pass the property test $\psi$ (i.e., reproduce the failure considered). \ProbDD \cite{probdd}, a recent advancement, enhances \CHISEL by substituting \ddmin with a probabilistic 
algorithm that increases each element's likelihood of successfully passing the property test $\psi$.

We use the original open-source implementations of both \CHISEL and \ProbDD, with \ProbDD adapted and embedded into \CHISEL's framework. For \PMA, we have integrated it with \CHISEL to skip redundant test cases and avoid unnecessary executions. Additionally, we will discuss why directly integrating \PMA with \ProbDD does not result in significant benefits (\Cref{sec:RQ2}).


This tripartite comparison allows us to comprehensively evaluate \PMA's advantages against leading representatives of both traditional and recent improvements in program reduction.

To ensure a fair and consistent evaluation of these three 
tools, we disabled features like Dead Code Elimination (DCE) and Dependency Analysis (DA). This decision aligns with findings from previous studies \cite{probdd}, which showed that these features could skew program reduction outcomes. Notably, DCE, while effective in debloating, can misfire in scenarios with compiler bugs triggered by unreachable code. Similarly, DA has been observed to produce erroneous results, especially when functions are passed as parameters.

\begin{table*}[htbp]
\centering
\caption{Comparison of \CHISEL, \ProbDD, and \PMA. For each subject, O(\#) shows the original token count, S(\#) the reduced token count, P(s) the processing time (in seconds), R(\#/s) the reduction speed  (tokens per second), and T(\#) the number of test cases executed. \PMA's improvements are shown relative to both \CHISEL and \ProbDD.
}
\vspace*{-1ex}
\label{tab:comparison}
\small
\scalebox{0.74}{
\begin{tabular}{
>{\columncolor[HTML]{FFFFFF}}c 
>{\columncolor[HTML]{FFFFFF}}c 
>{\columncolor[HTML]{CFCFCF}}c 
>{\columncolor[HTML]{DFDFDF}}c 
>{\columncolor[HTML]{EFEFEF}}c 
>{\columncolor[HTML]{F7F7F7}}c
>{\columncolor[HTML]{CFCFCF}}c 
>{\columncolor[HTML]{DFDFDF}}c 
>{\columncolor[HTML]{EFEFEF}}c 
>{\columncolor[HTML]{F7F7F7}}c
>{\columncolor[HTML]{CFCFCF}}c 
>{\columncolor[HTML]{DFDFDF}}c 
>{\columncolor[HTML]{EFEFEF}}c 
>{\columncolor[HTML]{F7F7F7}}c
>{\columncolor[HTML]{CFE8F6}}c 
>{\columncolor[HTML]{DFF2FA}}c 
>{\columncolor[HTML]{EFF7FD}}c 
>{\columncolor[HTML]{F7FBFE}}c
>{\columncolor[HTML]{CFE8F6}}c 
>{\columncolor[HTML]{DFF2FA}}c 
>{\columncolor[HTML]{EFF7FD}}c 
>{\columncolor[HTML]{F7FBFE}}c}

\toprule
\multicolumn{1}{l}{\cellcolor[HTML]{FFFFFF}} &
\multicolumn{1}{l}{\cellcolor[HTML]{FFFFFF}} &
\multicolumn{4}{c}{\CHISEL} &
\multicolumn{4}{c}{\ProbDD} &
\multicolumn{4}{c}{\PMA} &
\multicolumn{4}{c}{\PMA vs. \CHISEL} &
\multicolumn{4}{c}{\PMA vs. \ProbDD} \\

\cmidrule(lr){3-6} \cmidrule(lr){7-10} \cmidrule(lr){11-14} \cmidrule(lr){15-18} \cmidrule(lr){19-22}
\multirow{-2}{*}{\cellcolor[HTML]{FFFFFF}Subject} &
\multirow{-2}{*}{\cellcolor[HTML]{FFFFFF}O(\#)} &
S(\#) & P(s) & R(\#/s) & T(\#) &
S(\#) & P(s) & R(\#/s) & T(\#) &
S(\#) & P(s) & R(\#/s) & T(\#) &
{↑S(\%)} & {↑P(\%)} & {×R} & {↑T(\%)} &
{↑S(\%)} & {↑P(\%)} & {×R} & {↑T(\%)} \\
\midrule
mkdir-5.2.1 & 34801 & 8741 & 3517.7 & 7.41 & 19452 & 8765 & 4340.5 & 6.00 & 14265 & 8817 & 3007.1 & 8.64 & 12041 & -0.9\% & 14.5\% & 1.17 & 38.1\% & -0.6\% & 30.7\% & 1.44 & 15.6\% \\
rm-8.4 & 44459 & 8027 & 8888.4 & 4.10 & 26274 & 8213 & 4624.6 & 7.84 & 5171 & 8240 & 4019.4 & 9.01 & 7110 & -2.7\% & 54.8\% & 2.20 & 72.9\% & -0.3\% & 13.1\% & 1.15 & -37.5\% \\
chown-8.2 & 43869 & 9212 & 9428.8 & 3.68 & 24523 & 9116 & 7476.5 & 4.65 & 5534 & 9214 & 4574.5 & 7.58 & 5466 & 0.0\% & 51.5\% & 2.06 & 77.7\% & -1.1\% & 38.8\% & 1.63 & 1.2\% \\
grep-2.19 & 127681 & 103549 & - & 2.23 & 15093 & 99331 & - & 2.63 & 2354 & 91056 & - & 3.39 & 2999 & 12.1\% & 0.0\% & 1.52 & 80.1\% & 8.3\% & 0.0\% & 1.29 & -27.4\% \\
bzip2-1.05 & 70530 & 65173 & - & 0.50 & 17867 & 43102 & - & 2.54 & 5112 & 39906 & - & 2.84 & 5912 & 38.8\% & 0.0\% & 5.72 & 66.9\% & 7.4\% & 0.0\% & 1.12 & -15.6\% \\
sort-8.16 & 88068 & 54541 & - & 3.10 & 16273 & 57876 & - & 2.80 & 1211 & 42795 & - & 4.19 & 1947 & 21.5\% & 0.0\% & 1.35 & 88.0\% & 26.1\% & 0.0\% & 1.50 & -60.8\% \\
gzip-1.2.4 & 45929 & 30970 & - & 1.39 & 12638 & 28333 & - & 1.63 & 3611 & 30283 & - & 1.45 & 3402 & 2.2\% & 0.0\% & 1.05 & 73.1\% & -6.9\% & 0.0\% & 0.89 & 5.8\% \\
uniq-8.16 & 63861 & 14399 & 8925.1 & 5.54 & 14736 & 14400 & 8765.7 & 5.64 & 3974 & 14243 & 6654.8 & 7.46 & 4090 & 1.1\% & 25.4\% & 1.35 & 72.2\% & 1.1\% & 24.1\% & 1.32 & -2.9\% \\
date-8.21 & 53442 & 38861 & - & 1.35 & 40953 & 22320 & - & 2.88 & 11846 & 20934 & - & 3.01 & 11268 & 46.1\% & 0.0\% & 2.23 & 72.5\% & 6.2\% & 0.0\% & 1.04 & 4.9\% \\
tar-1.14 & 163296 & 157200 & - & 0.56 & 25155 & 89572 & - & 6.83 & 3551 & 27982 & - & 12.53 & 5251 & 82.2\% & 0.0\% & 22.20 & 79.1\% & 68.8\% & 0.0\% & 1.84 & -47.9\% \\
\midrule
clang-22382 & 9987 & 3416 & 554.1 & 11.86 & 4661 & 3416 & 176.0 & 37.34 & 558 & 3416 & 139.7 & 47.04 & 705 & 0.0\% & 74.8\% & 3.97 & 84.9\% & 0.0\% & 20.6\% & 1.26 & -26.3\% \\
clang-22704 & 184444 & 1333 & 5605.8 & 32.66 & 8759 & 1063 & 6100.6 & 30.06 & 1027 & 1333 & 3496.2 & 52.37 & 1016 & 0.0\% & 37.6\% & 1.60 & 88.4\% & -25.4\% & 42.7\% & 1.74 & 1.1\% \\
clang-23309 & 33310 & 7116 & 3116.0 & 8.41 & 16660 & 7111 & 1300.2 & 20.15 & 1558 & 7116 & 982.9 & 26.65 & 1896 & 0.0\% & 68.5\% & 3.17 & 88.6\% & -0.1\% & 24.4\% & 1.32 & -21.7\% \\
clang-23353 & 30196 & 7741 & 2347.0 & 9.57 & 14829 & 7698 & 632.0 & 35.60 & 1179 & 7741 & 573.8 & 39.13 & 1708 & 0.0\% & 75.6\% & 4.09 & 88.5\% & -0.6\% & 9.2\% & 1.10 & -44.9\% \\
clang-25900 & 78960 & 3016 & 2820.7 & 26.92 & 11032 & 2988 & 1116.4 & 68.05 & 914 & 3016 & 962.5 & 78.90 & 1188 & 0.0\% & 65.9\% & 2.93 & 89.2\% & -0.9\% & 13.8\% & 1.16 & -30.0\% \\
clang-26760 & 209577 & 5241 & 4111.1 & 49.70 & 6484 & 5241 & 2238.7 & 91.27 & 959 & 5241 & 2147.0 & 95.17 & 1050 & 0.0\% & 47.8\% & 1.91 & 83.8\% & 0.0\% & 4.1\% & 1.04 & -9.5\% \\
clang-27137 & 174538 & 13002 & 5938.5 & 27.20 & 14576 & 14106 & 2403.4 & 66.75 & 2496 & 13176 & 2168.8 & 74.40 & 2625 & -1.3\% & 63.5\% & 2.74 & 82.0\% & 6.6\% & 9.8\% & 1.11 & -5.2\% \\
clang-27747 & 173840 & 6355 & 995.2 & 168.29 & 5763 & 6340 & 684.6 & 244.67 & 624 & 6355 & 387.7 & 432.00 & 634 & 0.0\% & 61.0\% & 2.57 & 89.0\% & -0.2\% & 43.4\% & 1.77 & -1.6\% \\
clang-31259 & 48799 & 4166 & 1449.7 & 30.79 & 7273 & 4166 & 610.9 & 73.06 & 1001 & 4166 & 513.0 & 87.00 & 1272 & 0.0\% & 64.6\% & 2.83 & 82.5\% & 0.0\% & 16.0\% & 1.19 & -27.1\% \\
gcc-59903 & 57581 & 6657 & 1900.9 & 26.79 & 9816 & 6657 & 832.4 & 61.18 & 1336 & 6657 & 771.1 & 66.04 & 1575 & 0.0\% & 59.4\% & 2.47 & 84.0\% & 0.0\% & 7.4\% & 1.08 & -17.9\% \\
gcc-60116 & 75224 & 9206 & 2210.3 & 29.87 & 12630 & 9188 & 998.1 & 66.16 & 1643 & 9206 & 685.1 & 96.36 & 1920 & 0.0\% & 69.0\% & 3.23 & 84.8\% & -0.2\% & 31.4\% & 1.46 & -16.9\% \\
gcc-61383 & 32449 & 9581 & 1697.9 & 13.47 & 10564 & 9581 & 1054.6 & 21.68 & 1933 & 9581 & 478.5 & 47.79 & 1598 & 0.0\% & 71.8\% & 3.55 & 84.9\% & 0.0\% & 54.6\% & 2.20 & 17.3\% \\
gcc-61917 & 85359 & 13064 & 2003.3 & 36.09 & 10289 & 13064 & 999.7 & 72.32 & 1388 & 13064 & 639.3 & 113.08 & 1339 & 0.0\% & 68.1\% & 3.13 & 87.0\% & 0.0\% & 36.1\% & 1.56 & 3.5\% \\
gcc-64990 & 148931 & 7347 & 2535.9 & 55.83 & 10231 & 7195 & 1004.9 & 141.04 & 1425 & 7347 & 865.7 & 163.55 & 1578 & 0.0\% & 65.9\% & 2.93 & 84.6\% & -2.1\% & 13.9\% & 1.16 & -10.7\% \\
gcc-65383 & 43942 & 4969 & 1849.1 & 21.08 & 12750 & 4969 & 439.2 & 88.74 & 1160 & 4969 & 404.0 & 96.47 & 1423 & 0.0\% & 78.2\% & 4.58 & 88.8\% & 0.0\% & 8.0\% & 1.09 & -22.7\% \\
gcc-66186 & 47481 & 6444 & 1466.5 & 27.98 & 6542 & 6444 & 730.0 & 56.22 & 1088 & 6444 & 688.3 & 59.62 & 1238 & 0.0\% & 53.1\% & 2.13 & 81.1\% & 0.0\% & 5.7\% & 1.06 & -13.8\% \\
gcc-66375 & 65488 & 5020 & 3430.5 & 17.63 & 15657 & 5020 & 1051.5 & 57.51 & 1410 & 5020 & 900.3 & 67.16 & 1509 & 0.0\% & 73.8\% & 3.81 & 90.4\% & 0.0\% & 14.4\% & 1.17 & -7.0\% \\
gcc-70127 & 154816 & 12452 & 2535.8 & 56.14 & 9415 & 13000 & 1174.2 & 120.78 & 1426 & 12457 & 1058.4 & 134.50 & 1438 & 0.0\% & 58.3\% & 2.40 & 84.7\% & 4.2\% & 9.9\% & 1.11 & -0.8\% \\
gcc-70586 & 212259 & 8383 & 2291.3 & 88.98 & 7271 & 8383 & 893.1 & 228.28 & 748 & 8335 & 903.4 & 225.73 & 994 & 0.6\% & 60.6\% & 2.54 & 86.3\% & 0.6\% & -1.2\% & 0.99 & -32.9\% \\
gcc-71626 & 4397 & 639 & 12.1 & 310.58 & 184 & 639 & 12.4 & 303.06 & 123 & 639 & 5.2 & 722.69 & 70 & 0.0\% & 57.0\% & 2.33 & 62.0\% & 0.0\% & 58.1\% & 2.38 & 43.1\% \\
\midrule
mean & 86917 & 20861 & 3318.0 & 35.99 & 13612 & 17243 & 2069.2 & 64.24 & 2688 & 14292 & 1913.1 & 92.86 & 2875 & 6.7\% & 59.2\% & 3.32 & 80.5\% & 3.0\% & 22.0\% & 1.34 & -13.0\% \\
\bottomrule
\end{tabular}
}
\vspace*{-2ex}
\end{table*}

\subsection{Experiment Setup}

\subsubsection{Subjects}

To ensure robust evaluation and minimize selection bias, our study employs established benchmarks recognized in prior research, comprising two datasets:

\begin{itemize}
    \item \textbf{CHISEL Benchmark Suite:} This dataset, employed in \CHISEL \cite{chisel} and \ProbDD \cite{probdd} evaluations, comprises 10 C programs from GNU packages. These are typically used in embedded systems after debloating, feature complex property test functions $\psi$ involving numerous properties.

    \item \textbf{Perses Benchmark Suite:} Adopted from the Perses benchmark suite \cite{perses}, extensively used in delta debugging research including ProbDD \cite{probdd}, T-PDD \cite{tpdd}, and Vulcan \cite{vulcan}, this dataset features 20 C programs that trigger crashes or miscompilation issues 
    in various GCC (4.8.2, 4.9.0, 4.8.0) and LLVM (3.6.0, 3.7.0, 3.8.0) versions.  
\end{itemize}




\subsubsection{Metrics}

Our evaluation utilizes the following metrics:
\begin{itemize}
    \item \textbf{Reduced Program Size (\emph{S(\#)}):} The number of tokens in the reduced program.
    \item \textbf{Processing Time (\emph{P(s)}):} The total time taken for the reduction process, in seconds.
    \item \textbf{Test Case Count (\emph{T(\#)}):} The number of test cases actually executed.
    \item \textbf{Reduction Speed (\emph{R(\#/s)}):} The token deletion rate per second.
    \item \textbf{Performance Improvement (\(\uparrow\!\! M(\%)\)):} The percentage improvement of a metric $M\in \{S, P, T\}$ with \PMA over a baseline:
    \(
    \uparrow M(\%) = \left( \frac{M_{\text{baseline}} - M_{\PMA}}{M_{\text{baseline}}} \right) \times 100\%
    \)
    \item \textbf{Speedup Factor (\(\times R\)):} The ratio of reduction speeds of \PMA to a baseline:
    \(
    \times R = \frac{R_{\PMA}}{R_{\text{baseline}}}
    \)

\end{itemize}


For each 
tool, we documented the original size of each subject, the size of the smallest failure-inducing program found, the number of test cases executed, and the processing time. We also measured efficiency by calculating the token deletion rate per second, dividing the total tokens reduced by the total processing time.

To ensure fairness and handle potential timeouts, we applied a 3-hour timeout for each reduction process, following \cite{probdd, deepdiveprobdd}. 
In timeout cases, we recorded the smallest failure-inducing test case and marked the processing time as '-' to indicate the time limit was reached. The token deletion rate was still calculated, but timeout cases were excluded from percentage reduction in processing time calculations $\uparrow P(\%)$ across the 30 subjects.

\subsubsection{Computing Platform}

All experiments were conducted on a Linux server featuring an 8-core, 16-thread Intel(R) Xeon(R) CPU E5-1660 v4 @ 3.20 GHz, 256 GB of RAM, and running Ubuntu 16.04.


\subsection{RQ1: \CHISEL vs. \PMA}
\label{sec:RQ1}

As detailed in Table~\ref{tab:comparison}, \PMA significantly enhances \CHISEL's efficiency without compromising its effectiveness. Across 30 test subjects, on average, \PMA cuts processing time by 59.2\%, increases the token deletion rate by $3.32\times$, and reduces the size of the final reduced programs by 6.7\%. 

This improved performance is due to \PMA avoiding numerous redundant tests that \CHISEL executes, with percentage reductions ($\uparrow\! \text{T(\%)}$) from 38.1\% (\texttt{mkdir-5.2.1}) to 90.4\% (\texttt{gcc-66375}), averaging 80.5\%. Consequently, \PMA consistently achieves faster token deletion rates, ranging from $1.05\times$ (\texttt{gzip-1.2.4}) to $22.20\times$ (\texttt{tar-1.15}), and reduces processing time in all non-timeout cases ($\uparrow \text{P(\%)} > 0$), from $14.5\%$ (\texttt{mkdir-5.2.1}) to $78.2\%$ (\texttt{gcc-65383}).

\PMA achieves efficiency gains over \CHISEL while maintaining or improving effectiveness. For three out of 30 subjects—\texttt{mkdir-5.2.1}, \texttt{rm-8.4}, and \texttt{clang-27137}—\PMA slightly increased the sizes of the reduced programs compared to \CHISEL (\(\uparrow\!\text{S(\%)} = -0.9\%\), \(\uparrow\!\text{S(\%)} = -2.7\%\), and \(\uparrow\!\text{S(\%)} = -1.3\%\)), but reduced processing times by 14.5\%, 54.8\%, and 63.5\%, respectively. For \texttt{uniq-8.16}, the best among the non-timeout cases, \PMA reduced both program size by 1.1\% and processing time by 25.4\%.
Both \CHISEL and \PMA encountered timeouts on six programs, including \texttt{grep-2.19}, \texttt{bzip2-1.0.5}, \texttt{sort-8.16}, \texttt{gzip-1.2.4}, \texttt{date-8.21}, and \texttt{tar-1.14}, due to complex property test functions $\psi$ (used in program debloating). 
By eliminating redundant tests, \PMA sped up token deletion and produced smaller programs than \CHISEL, demonstrating its efficiency gains even when occasionally resulting in slightly larger reduced programs.

However, it should be noted that \PMA's use of a probabilistic model introduces subtle variations in specific cases. This approach, which allows for selective skipping of test cases, can result in minor divergences in results. For instance, we observed such variations in program size reduction rates ($\uparrow S(\%)$) for the 20 Perses benchmarks in \Cref{tab:comparison}, as reported by \PMA compared to \CHISEL.

\PMA and \CHISEL each produced 19 identically sized reduced programs (\(\uparrow\! S(\%)=0\)), with one from the \CHISEL benchmark suite \cite{chisel} and 18 from the Perses Benchmark Suite \cite{perses}. Excluding \texttt{chown-8.2}, both tools generated the same final reduced program for each of the remaining 18 programs, highlighting \PMA's enhancements in processing time and token deletion rates, alongside maintaining comparable effectiveness to \CHISEL.

\begin{figure*}[h]
    \centering
    \vspace*{0ex}
    \begin{subfigure}{1\linewidth}
        \centering        
        \includegraphics[width=\textwidth,height=0.12\textheight]{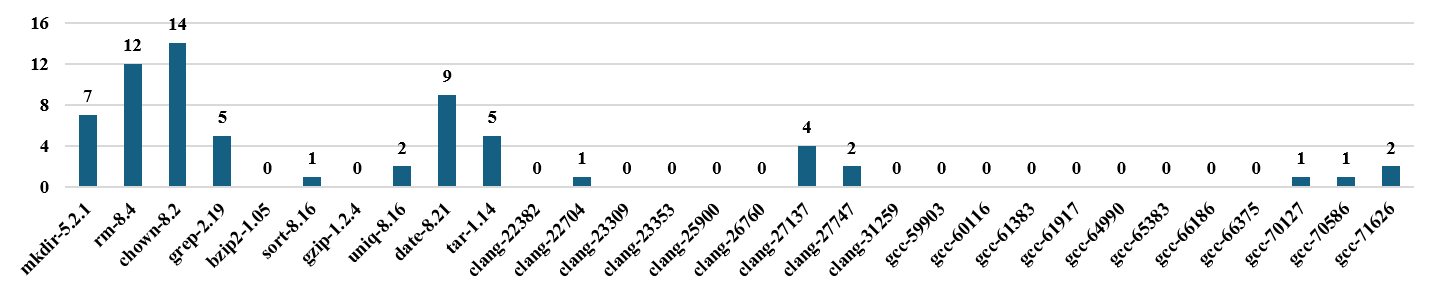}
        \caption{Number of failure-inducing test cases eliminated as being redundant by \PMA.}
        \label{fig:CHISEL-truecases}
    \end{subfigure}   
    \hfill       
    \vspace*{-2.5ex}
    \begin{subfigure}{1.0\linewidth}
        \centering
        \includegraphics[width=\textwidth,height=0.13\textheight]{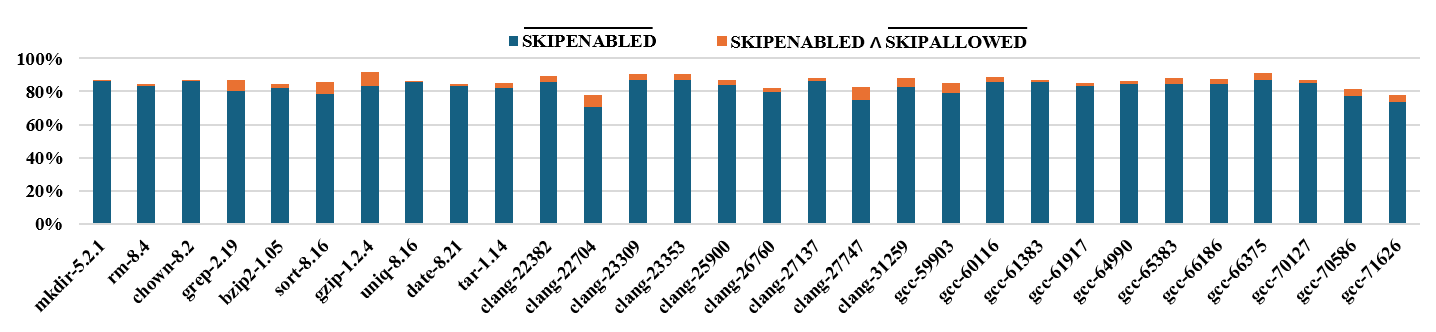}
        \caption{Percentage of non-failure-inducing test cases executed under \PMA, categorized by their compliance with $\overline{\monoCheck} \vee \overline{\skipTest}$ (line B5 of \Cref{fig:model}). Each bar represents a combination of test cases: those failing the monotonicity check ($\overline{\monoCheck}$) shown in blue, and those passing the monotonicity check but failing the skip test ($\monoCheck \wedge \overline{\skipTest}$) shown in orange.
        }
        \label{fig:CHISEL-falsecases}
    \end{subfigure}
     \vspace*{-5ex}
    \caption{Nature of test cases eliminated by \PMA.}
    \label{fig:CHISEL-truefalsecases}
    \Description{Nature of redundant test cases eliminated by \PMA.}
    \vspace*{-.5ex}
\end{figure*}

To further explore the findings presented in \Cref{tab:comparison}, \Cref{fig:CHISEL-truefalsecases} illustrates \PMA's effectiveness in selectively skipping tests that \CHISEL executes, while also offering insights into potential improvements by highlighting its limitations. 

\Cref{fig:CHISEL-truefalsecases}(a) illustrates the number of failure-inducing test cases \PMA eliminates for each program, complementing the totals in \Cref{tab:comparison}. While \PMA effectively reduces program size by skipping non-failure-inducing tests, it sometimes inadvertently skips failure-inducing tests due to violations of monotonicity (Assumption~\ref{def:monotonicity}) and inconsistency (Assumption~\ref{def:consistency}), posing two distinct challenges to delta debugging. 
While our approach generally improves efficiency by reducing test executions, it risks missing further reduction opportunities when tests are incorrectly omitted.

\begin{sloppypar}
Among the 30 subjects in \Cref{tab:comparison}, \PMA produced larger reduced programs for three subjects: \texttt{mkdir-5.2.1}, \texttt{rm-8.4}, and \texttt{clang-27137}. Focusing on \texttt{clang-27137}, improper optimization in \texttt{clang-3.8} under the ``\texttt{-O3}" setting triggers a bug due to an always-false loop, resulting in erroneous output. As shown in \Cref{fig:CHISEL-truefalsecases}(a), \PMA inadvertently skipped four failure-inducing tests for \texttt{clang-27137}, yielding a reduced program 1.3\% larger than \CHISEL's. We denote the \(N\)-th test sequence in \CHISEL for \texttt{clang-27137} as \(T_N\) (starting from 1). The skipping of these four tests—\(T_{3095}\), \(T_{4012}\), \(T_{4120}\), and \(T_{7922}\)—highlights two key challenges in \PMA's reduction strategy.
\end{sloppypar}

The first skipped test, \(T_{3095}\), illustrates the challenge of handling inconsistency in program reduction. \PMA inadvertently skipped it because \(\monoCheck(T_{3095}) = \true\) based on the superset \(T_{3091}\), which failed with a "\texttt{label used but not defined}" error, making $\psi(T_{3091}) = \false$. Removing the label in \(T_{3091}\) produced \(T_{3095}\), which fixed the error and reproduced the bug, rendering $\psi(T_{3095}) = \true$. Though this appears to violate monotonicity (Assumption~\ref{def:monotonicity}), it actually reflects a violation of consistency (Assumption~\ref{def:consistency}), as $\psi(T_{3091}) \notin \{\true, \false\}$. This case highlights the challenge of accurate skip decisions when consistency is violated.


\begin{sloppypar}
Recent advances in delta debugging have focused on consistent sequences from a given input program \cite{hdd,creduce,perses}, yet cases like \(T_{3091}\) may still yield inconsistent results. A potential solution is to refine the categorization of non-failure-inducing tests by expanding the `\false' category to include $\{\false, \texttt{unknown}\}$ \cite{zeller1999} (Assumption~\ref{def:consistency}), or even further, to $\{\false, \texttt{unknown}_1, \texttt{unknown}_2, ..., \texttt{unknown}_n\}$, which could strengthen Assumption~\ref{def:monotonicity} for more accurate test elimination.
\end{sloppypar}

Studying such \texttt{unknown} cases and their monotonicity patterns is crucial.
Failures from a missing main function often persist in subsets, while errors like undefined variables may resolve when problematic usage is removed. By categorizing and analyzing these cases, we can develop heuristics to better predict test outcomes and refine delta debugging strategies. We hope this work offers valuable insights for improving our approach in this direction.

The remaining three failure-inducing test cases—\(T_{4012}\), \(T_{4120}\), and \(T_{7922}\) in \texttt{clang-27137}—pose a distinct challenge. \PMA skipped these because \(\psi\) evaluated their corresponding supersets—\(T_{4003}\), \(T_{4116}\), and \(T_{7911}\)—as \false, even though these supersets executed correctly while their subsets triggered the bug. This discrepancy occurred when removing if statements from the supersets allowed for more aggressive optimizations, thus exposing the latent bug. These cases reveal violations of the monotonicity assumption critical to \PMA's skip mechanism, highlighting missed reduction opportunities. They also emphasize the need for further research into monotonicity violations, which may differ across various programs and delta debugging contexts.

Among the four failure-inducing test cases skipped by \PMA in \texttt{clang-27137}, skipping \(T_{7922}\) had the most significant impact on reduction efficacy. This caused \PMA to miss the opportunity to remove 174 additional tokens, resulting in \(\uparrow S(\%) = 1.3\%\) (\Cref{tab:comparison}).

\begin{sloppypar}
Let us examine \Cref{fig:CHISEL-truefalsecases}(b), which shows a stacked bar graph of non-failure-inducing test cases \(T\) executed under \PMA. Each bar is categorized as (1) \(\monoCheck(T)=\false\) (\textcolor{blue}{blue}), (2) \(\monoCheck(T)=\true \wedge \skipTest(T) = \false\) (\textcolor{orange}{orange}), and (3) the sum of categories (1) and (2) representing the overall bar height. These categories reflect the proportion of each type relative to the total executed by \PMA. Test cases in Category (1) are not skipped due to no prior non-failure-inducing supersets (Assumption~\ref{def:monotonicity}), while those in Category (2) persist based on the probabilistic model (\Cref{eq:skiptest}), with specific percentages ranging from 0.18\% (\texttt{mkdir-5.2.1}) to 8.20\% (\texttt{gzip-1.2.4}), averaging 3.46\%. 
\end{sloppypar}
This data highlights the need for advanced techniques to preemptively suppress tests in Category (1). It also supports our strategy of not using costly set-equality checks to completely eliminate all non-failure-inducing tests (e.g., those marked by $\texttt{F}^*$ in \Cref{tab:PMA-process}) that were executed earlier (\Cref{sec:modelapp}). It is justified because such tests are mostly skipped; those not successfully skipped would fall into Category (2), which are minimal in number.

In summary, \PMA significantly improves \CHISEL's efficiency while maintaining or enhancing its effectiveness, despite occasionally skipping some failure-inducing test cases.

\subsection{RQ2: \ProbDD vs. \PMA}
\label{sec:RQ2}

As shown in Table~\ref{tab:comparison}, comparing \PMA with \ProbDD across 30 test subjects, on average, \PMA reduces processing time by 22.0\%, boosts the token deletion rate by $1.34\times$, and shrinks final program sizes by 3.0\%. \PMA outperforms \ProbDD in processing time for all non-timeout subjects except \texttt{gcc-70586} and exceeds \ProbDD in token deletion rate for all but \texttt{gzip-1.2.4} and \texttt{gcc-70586}. Notable gains include \texttt{chown-8.2} with a 38.8\% time reduction and $1.63\times$ increase in token deletion, and \texttt{gcc-61383} with a 54.6\% time reduction and $2.20\times$ increase in token deletion.

\begin{figure*}[ht]
    \centering
    \includegraphics[width=\textwidth,height=0.12\textheight]{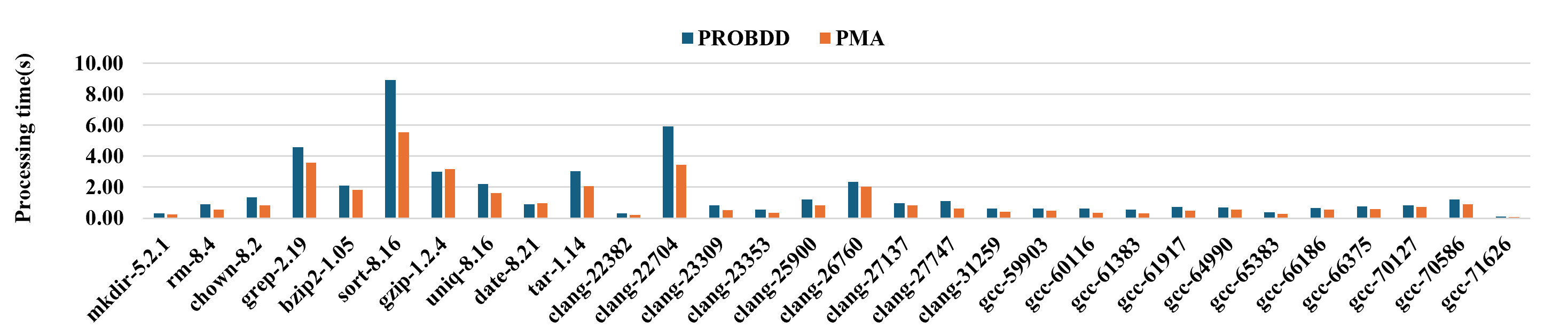}
    \vspace*{-5ex}
    \caption{Comparison of \ProbDD and \PMA in processing time
    per test case (with a 3-hour timeout).}
    \label{fig:ProbDD-processingtime}
        \vspace*{-1.75ex}
\end{figure*}

The performance differences between \PMA and \ProbDD are rooted in their distinct algorithmic approaches. \ProbDD replaces \CHISEL’s \ddmin component, using probabilistic analysis to better detect failure-inducing elements. Conversely, \PMA enhances \ddmin within \CHISEL, improving efficiency by employing a simple yet effective model to skip redundant tests.
Due to \ProbDD's focus on individual elements, it is more computationally intensive than \PMA's strategy, which manages entire test sequences. As illustrated in \Cref{fig:ProbDD-processingtime}, \PMA achieves an average reduction in processing time per test case of 26.4\% across 30 subjects. Excluding \texttt{gzip-1.2.4} and \texttt{date-8.21}, which show decreases of $-6.1\%$ and $-5.1\%$, respectively, the reductions for the remaining 28 subjects range from 10.6\% (\texttt{gcc-70127}) to 45.1\% (\texttt{gcc-61383}), averaging 28.7\%.

\begin{sloppypar}
Looking at program reduction effectiveness, \PMA generally outperforms \ProbDD, achieving an average decrease of 3.0\%. This improvement is particularly pronounced in cases like \texttt{clang-27137}, which sees a 6.6\% reduction. However, in instances such as \texttt{clang-22704}, while \PMA increases the program size by 25.4\%, it also cuts processing time by 42.7\%, highlighting the trade-off between efficiency and effectiveness. Judiciously navigating such trade-offs remains a challenging problem.
\end{sloppypar}

Comparing the token deletion rates, \PMA outperforms \ProbDD with an average speedup of \(1.34\times\). Notable improvements are seen in \texttt{gcc-71626}, achieving a speedup of \(2.38\times\). Conversely, \PMA slightly lags in \texttt{gcc-70586}, with a speedup of \(0.99\times\) and a minimal increase in processing time of 1.2\%, albeit with a slight improvement in program size reduction by 0.6\%. In all six programs that reached the 3-hour timeout limit, except for \texttt{gzip-1.2.4}, \PMA consistently produced smaller reduced programs at faster token deletion rates.

Integrating our \PMA approach directly into \ProbDD yielded negligible efficiency gains, highlighting both the challenges and opportunities of merging these strategies. The limited improvement stems from \ProbDD’s architecture within \CHISEL, where it replaces \ddmin and prioritizes smaller test cases to probabilistically detect failure-inducing elements. This strategy often generates non-failure-inducing cases, progressing from smaller to larger ones, diminishing the effectiveness of \PMA's skip mechanism. \PMA is crafted to eliminate redundant tests through structured, sequence-based reductions using a divide-and-conquer approach, similar to those seen in \ddmin-style algorithms.


Future research could explore combining \PMA's sequence-based reduction with \ProbDD's granular approach—using \PMA for rapid space reduction followed by \ProbDD's precise refinement. Alternatively, optimizing \ProbDD to prioritize efficiency, even if it introduces some redundancies, would give \PMA greater scope to eliminate excess tests, thereby enhancing overall efficiency.

In summary, our \PMA approach, combined with \CHISEL, significantly outperforms \ProbDD (also based on \CHISEL), further demonstrating its advancements in delta debugging.

\section{Related Work}
\label{sec:rel}

\noindent\textbf{\ddmin-based Delta Debugging.} 
The \texttt{ddmin} algorithm~\cite{ddmin} establishes a systematic foundation for test case reduction. It uses an iterative approach to minimize failure-inducing inputs, testing subsets to identify and isolate the minimal failure-inducing elements.

To overcome \ddmin's limitations with large, structural inputs, Hierarchical Delta Debugging (HDD)~\cite{hdd} extends \ddmin for hierarchical structures (e.g., parse trees), applying reduction root-to-leaf to handle large inputs efficiently. Later variants—Coarse HDD~\cite{coarsehdd}, modernized HDD~\cite{modernizinghdd}, and HDDr~\cite{hddr}—further optimize this approach.


\begin{sloppypar}


Perses~\cite{perses} advanced delta debugging by leveraging formal syntax to guide the reduction process, utilizing \ddmin to prune parse tree nodes while maintaining syntactic validity. 

\CHISEL~\cite{chisel}, a baseline for \PMA evaluation, employs reinforcement learning with a Markov decision model to prioritize \ddmin test cases more likely to reproduce failures.

\ProbDD~\cite{probdd}, another baseline for \PMA evaluation, improves \ddmin through Bayesian optimization, generating test cases based on the likelihood of each element’s success. T-PDD~\cite{tpdd} further refines this by leveraging syntactic relationships to estimate the probability of retaining each element in a test case.
\end{sloppypar}

Among existing probabilistic approaches, ProbDD and T-PDD use historical test data to update element probabilities and select high-probability test sequences. In contrast, \PMA focuses on search space monotonicity rather than individual element probabilities. 
While \CHISEL employs a statistical model to prioritize \ddmin sequences likely to pass the property test $\psi$, \PMA's monotonicity-based approach enables more flexible and efficient pruning of the search space. This distinctive strategy allows \PMA to potentially surpass these existing methods in both efficiency and effectiveness.
Our evaluation demonstrates that \PMA significantly improves performance compared to \CHISEL and \ProbDD, highlighting its potential to advance delta debugging techniques.


\begin{sloppypar}
\noindent\textbf{Combining \ddmin and Program Transformations.}
C-Reduce~\cite{creduce} focuses on deletion operations enhanced by language-specific transformations such as function inlining and constant propagation. Utilizing C/C++ semantics, it effectively reduces programs, often surpassing general-purpose reducers in creating minimal test cases that still expose bugs within the programs.
\end{sloppypar}

Similarly, ddSMT~\cite{ddsmt} is tailored for the SMT-LIBv2 format, using language-specific transformations like command-level scope and term substitution to effectively reduce SMT formulas. Its successor, ddSMT2.0~\cite{ddsmt2}, introduces a hybrid strategy combining SMT-specific transformations with \ddmin and hierarchical approaches while supporting all SMT-LIBv2 dialects.

Vulcan~\cite{vulcan} is a general framework that enhances \ddmin-based methods by integrating transformations. It starts with a \ddmin-based AGR (Agnostic Program Reduction), such as Perses \cite{perses}, then applies transformations like Identifier Replacement and Tree-Based Local Exhaustive Enumeration to create new reduction opportunities. By iteratively combining transformation and reduction, Vulcan surpasses traditional \ddmin's 1-minimality limitation.


Transformation-based approaches target smaller reductions but often face efficiency challenges. This paper enhances the efficiency of existing \ddmin-based methods, setting aside the refinement of transformation-based approaches for future research.

\section{Conclusion}
In this paper, we introduce Probabilistic Monotonicity Assessment (\PMA), a novel approach that enhances delta debugging by dynamically modeling search space monotonicity with probabilistic techniques. 
Unlike approaches that focus on individual element probabilities, \PMA assesses the entire search space, enabling more efficient pruning of test sequences. Through a dynamically updated confidence function, \PMA adapts to changes in monotonicity, significantly boosting efficiency by reducing the number of required tests while maintaining or enhancing effectiveness. This focus on search space monotonicity offers a fresh perspective on delta debugging strategies and suggests potential for future research to explore additional search space characteristics to further refine program reduction and bug localization in software engineering.

While \PMA enhances the efficiency and effectiveness of \ddmin-style algorithms such as \CHISEL~\cite{chisel}, opportunities for further improvement remain. First, although it reduces the number of tests executed by \CHISEL by 80.5\% on average (\Cref{tab:comparison}), \PMA achieves only a $3.32\times$ average speedup in token deletion rate, suggesting room to reduce analysis overhead for greater performance gains. Second, \PMA may inadvertently filter out failure-inducing tests, as illustrated in \Cref{fig:CHISEL-truefalsecases}(a), due to inconsistencies with Assumptions~\ref{def:monotonicity} and~\ref{def:consistency} in practice. Some subsets of a failing test case may still pass $\psi$. Expanding the categorization of non-failure-inducing tests—for example, broadening the \false category (in Assumption~\ref{def:consistency}) to $\{\false, \texttt{unknown}\}$~\cite{zeller1999}—could refine Assumption~\ref{def:monotonicity} and enable more nuanced elimination rules. Third, although \PMA significantly reduces the number of tests executed, many tests failing $\monoCheck$ are still retained (\Cref{fig:CHISEL-truefalsecases}(b)), suggesting the potential for further reduction techniques. Finally, while \PMA combined with \CHISEL outperforms \ProbDD, integrating \PMA directly with \ProbDD yields negligible gains, as detailed in \Cref{sec:RQ2}. \ProbDD reduces redundant tests from \ddmin but increases per-test processing time compared to \PMA-\CHISEL (\Cref{fig:ProbDD-processingtime}). Speeding up \ProbDD, even at the cost of some redundancy, could improve the \PMA-\ProbDD combination by giving \PMA more opportunities to eliminate excess tests. Alternatively, a hybrid approach could apply \PMA first to narrow the search space, followed by \ProbDD for finer-grained reduction.



\section{Acknowledgement}

This research is supported by 
an Australian Research
Council (ARC) Grant (DP210102409).

\section{Data Availability}
The artifacts and experimental data for this paper are publicly accessible at \url{https://doi.org/10.5281/zenodo.13756162}. 

\bibliographystyle{ACM-Reference-Format}
\bibliography{references}

\end{document}